# Debug Support, Calibration and Emulation for Multiple Processor and Powertrain Control SoCs

A. Mayer[1], H. Siebert[1], K.D. McDonald-Maier[2].
[1]Infineon Technologies AG, Automotive & Industrial, Munich, Germany.
[2]University of Kent, Department of Electronics, Canterbury, CT2 7NT, UK.

## Abstract

*The introduction of complex SoCs with multiple processor cores presents new development challenges, such that development support is now a decisive factor when choosing a System-on-Chip (SoC). The presented developments support strategy addresses the challenges using both architecture and technology approaches. The Multi-Core Debug Support (MCDS) architecture provides flexible triggering using cross triggers and a multiple core break and suspend switch. Temporal trace ordering is guaranteed down to cycle level by on-chip time stamping. The Package Sized-ICE (PSI) approach is a novel method of including trace buffers, overlay memories, processing resources and communication interfaces without changing device behavior. PSI requires no external emulation box, as the debug host interfaces directly with the SoC using a standard interface.*

## 1 Introduction

Strong consumer demand for more refined automatic gearboxes and engines, combined with tough new emissions standards drive increased requirements for power train controllers. Significant adoption of direct injection Diesel technology in Europe has also lead to further increases in real-time demands. Migration from system-on-Printed Circuit Board (PCB) to single processor core System-on-Chip (SoC) has so far supported the corresponding rise in system complexity. The new generations of SoCs, containing many active peripherals and now multiple processor cores are increasingly more difficult to program, especially for mission critical real-time systems such as automotive control.

Developers are increasingly overwhelmed by development challenges such that system reliability is decreasing. A recent study has shown that 77 percent of electronic failures in cars were due to software [1]. Detecting and correcting these bugs early in the product cycle before the customer discovers them prevents loss of reputation and customer loyalty, or even loss of life in safety critical applications. For complex systems like automotive power-train control it is of particular importance to understand and analyze the behaviour in all possible scenarios, in support high quality software and reliable products. With the right tools, developers can overcome the challenges, bringing a new world of exciting and dependable products on time and on budget. The development support's effectiveness is now a decisive factor when system developers choose between SoCs.

Rising development time and increased product failures provide clear motive for improved tools, as verification and development tasks after the first silicon is produced consume about half a system's development effort [2]. Furthermore the US national institute of standards and technology has estimated that bugs and glitches cost the US economy alone approximately $59.9 billion a year. Increased testing and more effective debugging could reduce this by a third [3].

System development has traditionally been aided by replacing production SoCs with special development devices or In-Circuit Emulators (ICE) [4, 5, 6]. An ICE is intended to have similar behaviour as the production SoC but provide increased debug support resources. One approach to constructing ICEs is the Bond-Out chip (BO), which is realised by wire bonding internal nodes to additional external device pins, thus making the internal nodes visible to bench equipment such as logic analysers. BO based emulation techniques have a number of significant disadvantages: 1) Internal system nodes in the BO the must drive extra connections which results in changed its behaviour from the production SoC. 2) SoCs targeted at harsh environments have external pins limited to frequencies lower than the internal circuits [2, 5], preventing production speed real-time operation. 3) BOs are large and fragile, preventing them being used for applications like gearbox control where the controller is mounted within the gearbox. 4) BOs have a different footprint to production devices, forcing the development of two different printed circuit board layouts, each with many high speed wires. 5) Construction of a BO requires a custom mask set which is already expensive, and the cost rises aggressively with increased integration. The fundamental flaw of BO devices is that systems using BO based ICEs behave significantly differently to production systems. This has led to a steep decline in BO based development systems.

The now more relevant alternative to dedicated ICEs, is on-chip debug support. The on-chip debug support circuits provide development infrastructure for run and control of processors by external debugging tools [6]. On-chip hardware breakpoint triggers allow developers to halt processor cores when a defined program address is reached, enabling 'post-mortem' debugging. Some cores support watchpoints to stop them when data locations are accessed. The on-chip debug support based ICE solution is widely used, as the ICE behavior closely matches the production SoC with debug support disabled. Post-mortem debugging has the disadvantage that developers are only able to get a good look at the system after it is halted, rather than while it is running

## 2 Mechanical systems

Mechanical systems require continuous control until they are safely shut down, which makes 'post-mortem' debugging impractical. Systems such as hard-disk drives and engines can be irreparably damaged if the controlling electronics are switched off or suddenly stopped by a processor's breakpoint. Many mechanical systems also require extensive calibration, i.e. tuning of control loop parameters after assembly and at run-time, making unobtrusive access to internal memories essential.

## 3 Multiple processors

Debugging systems with concurrency is seldom straightforward, but when multiple processors are unsynchronized development is especially challenging. Observation of shared variable accesses is critical to debugging such systems. Tracing with on-chip infrastructure to ensure temporal ordering of messages provides such observation, and fulfills the requirements for mechanical systems. Many high quality trace solutions already exist, such as the Nexus standard [4], however they fail to define the infrastructure required to guarantee the correct trace order.

For single processor core SoCs there is already a growing mismatch between circuit frequency and device pin frequency [2], making external trace interfaces less feasible. For two cores the trace data doubles making the problem worse. Furthermore developers only require key pieces of information not millions of cycles of unrelated trace.

The presented solution tackles the problem on two fronts. At the architecture level, complex triggers qualify or 'filter' the trace down to only the required messages. At the technology level, new techniques have been developed to store the trace within the device package and provide additional emulation resources.

## 4 Multi-core debug solution

The Multi-Core Debug Solution MCDS is a trigger and trace logic block including trace qualification and compression. The MCDS block shown by figure 1 is placed at each processor core, therefore trace from one or several cores can be recorded in parallel. Scalable time stamping not shown in the figures ensures that all messages are stored in correct temporal order. The time stamping allows a time resolution down to cycle level. For heterogeneous cores only the adaptation logic differs, making the solution sufficiently flexible for design reuse. The system centric approach supports tracing of on-chip multi-master buses and general system states, independently from the processor cores.

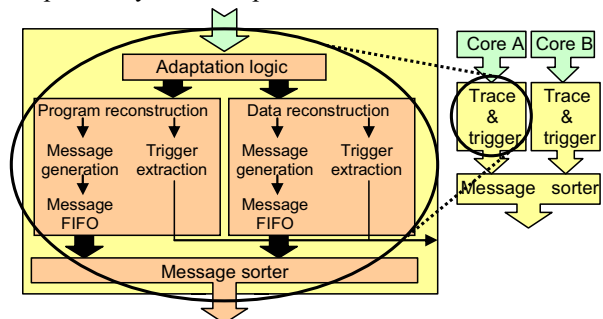

**Figure 1. MCDS trace and trigger block**

The trigger resources are implemented for the program and data accesses and are further enhanced using state-machines based on counters. They are compact but effective, especially with cross-triggering as shown by figure 2. Combining triggers from multiple sources is undefined by most previous solutions including the Nexus standard [4]. For example should a trigger stop one or multiple cores? The best solution is to let the developer decide by providing a reconfigurable break and suspend switch. When processor cores can be stopped the multi-core break and suspend switch is vital, as it halts synchronized cores without excessive slippage. The switch manages the response to both on-chip and external trigger inputs.

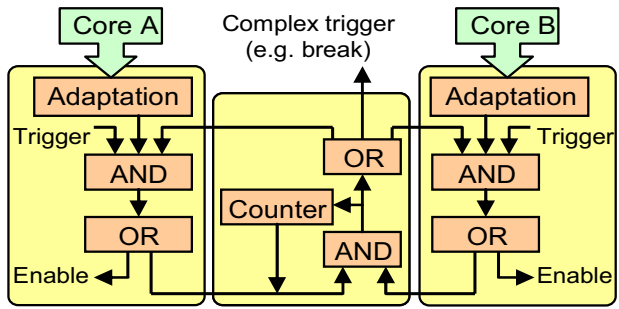

**Figure 2. Multiple core cross trigger unit**

## 5 Package sized ICE

The Package Sized In-circuit emulator (PSI) is a new approach to in-circuit emulation. With contemporary ICEs an external box contains some of the resources, however with PSI everything is included within the chip package. Integrating large calibration overlay memories and trace memories within a high volume SoC is not cost effective, as they are left idle in the final product. For high volume SoCs PSI overcomes these costs by moving the resources into a development specific part. Low volume SoCs do not justify the extra design effort for a development part, so the extra resources can be included in every chip as with conventional solutions. Crucially PSI uses the same footprint as the production SoC, a significant difference compared to bond-out based ICEs. A common foot print eliminates the vast effort of designing then debugging two versions of a system, including its PCB. A range of communication interfaces are possible with PSI, although Universal Serial Bus (USB) provides a straightforward connection to the debugging host computer, avoiding unwieldy proprietary interfaces. PSI allows full emulation support in the target system, even under extreme form factor constraints, allowing use within a gearbox. It is even possible to avoid a dedicated debugger interface altogether and use spare bandwidth on any accessible system interface.

A range of construction techniques have been developed for PSI, including both one and two chip versions. Maintaining consistent behavior between production and development SoCs is paramount, especially for analogue circuits, so the same manufacturing technology and flow are used for both development and production SoCs. For single chip PSI a development specific integrated circuit is created, where the entire production SoC layout, including bonding pads is treated as a hard macro. Single chip PSI is so far realized as an emulation side booster, which is a region located at the edge of the SoC macro as shown by figure 3. The main disadvantage of any solution requiring extra mask sets is that for future technologies cost may be prohibitive.

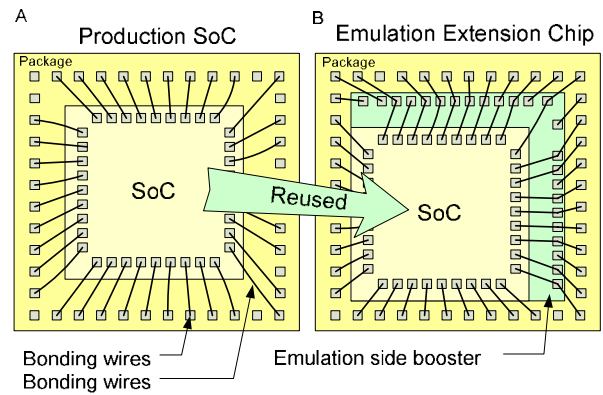

**Figure 3. Single chip emulation extension**

PSI with two chips adds extra development resources using an emulation extension chip. The emulation extension chip can either take the form of a carrier chip to which the production SoC is bonded, or as a booster chip attached to the production SoC. The extension chip PSI approaches are shown by figure 4. One advantage of the two chip approach is that the development specific chip could be reused across a product range.

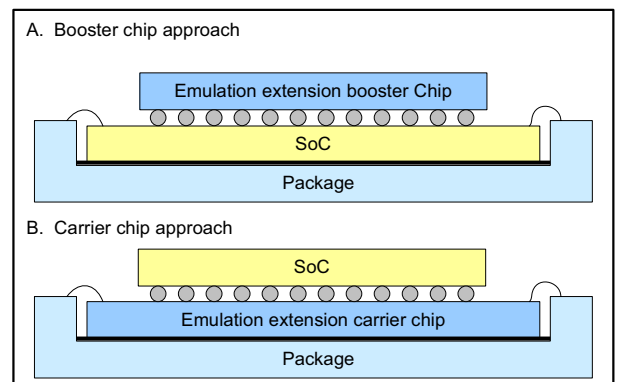

**Figure 4. Two chip emulation extension**

## 6 Implementation

The emulation side booster concept has been realized for the TC1796 power train controller SoC manufactured in 0.13µm technology. Figure 5 shows the packages used for the TC1796 production part and the corresponding TC1796 development part. The development SoC includes an extra 512Kbytes of SRAM, a USB 1.1 peripheral and a further processor core to service debug requests. Both versions of the SoC are interchangeable with complete transparency to the application system, while significantly boosting development support.

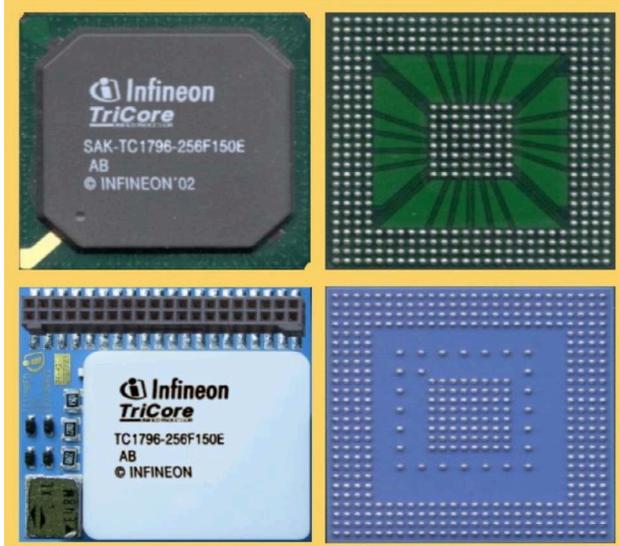

**Figure 5. Production and PSI devices**

The USB 1.1. interface has significant software overhead, but the system is unaffected as an extra PCP2 processor core is integrated to run the supplied driver. The extra processor can also be used for performance monitoring and consistency checking, and provides a new programmable tool not found in previous ICEs. Furthermore a robust calibration system is implemented using the universal measurement and calibration protocol XCP [7] over USB, or for extreme form factors an existing CAN interface. The only equipment needed besides the development SoC is a galvanically decoupled USB cable and the supplied XCP driver. For control actions requiring low latency the JTAG based interface's 2µs latency is more suitable than the 3ms of the USB interface. Cold booting of the extra processor is supported by having a separate power connection for the emulation memory.

## 7 Application development and calibration

The TC1796ED provides greatly increased flexibility over conventional parts due to its significant extra resources and configurability. During testing developers found using the 512kByte emulation RAM to hold the program highly beneficial for initial development. Not only does this avoid continuous reprogramming of the large 2 MByte program flash memory, but unlimited software breakpoints are possible, as with development of desktop applications. The emulation RAM is segmented into 64 kByte blocks for use as either overlay or trace memory. An address-mapping block resides on the production chip. It allows memory access redirection for up to 16 address ranges, with individual block sizes from 1 kByte to 32 kBytes of the overlay Emulation RAM. The access timing matches the flash memory being overlaid, ensuring consistent behavior. The overlay memory is divided into two pages that can be swapped atomically by a single control access. The large size of the RAM is determined by the automotive specific flash overlay requirements for calibration. The trace features used for system debug of mission critical real-time systems require just a fraction of that. The on-chip time stamping maintains temporal message ordering for all program and data flows including the full data activity of the multi-master busses.

## 8 Conclusion

The combination of MCDS and PSI provide the features developers need to manage the rising challenges of using complex SoCs within complex systems. The functionality of PSI scales well with technology advancement as it uses the same manufacturing flow as the production SoC. The main limitation of current PSI solutions is that extra mask sets will be more expensive in the future, but not prohibitive for a few more generations. There is however a smooth path then to integrate PSI part on each production SoC in particular for the case when no large calibration overlay memory is required. Selective integration of the emulation side booster using only a single set of masks is one alternative solution. Selective integration of PSI is an area of active research, and has been possible for a small region on one side of the SoC. No matter which implementation used, PSI with MCDS boosts development support firmly into the multiple processor core era.

## Acknowledgements

A.B.T. Hopkins provided valuable support and feedback during the writing of this paper. The research of the author K.D. McDonald-Maier is supported in part through EPSRC grant GR/S13361/01.

## References

[1] G. Jacobi, "Software ist im Auto ein Knackpunkt", VDI Nachrichten, 28th February 2003.
[2] Semiconductor Industry Association (SIA): International Technology Roadmap for Semiconductors, http://public.itrs.net/
[3] U.S. Department of Commerce, The economic impacts of inadequate infrastructure for software testing, tech. report RTI-7007.011US, National institute of standards and technology, US, May 2002.
[4] The Nexus 5001 Forum™, Standard for a global embedded processor debug interface, IEEE-ISTO 5001™-1999, http://www.nexus5001.org.
[5] A. Mayer, H. Siebert, A. Kolof, and S. el Baradie, "Debug Support for Complex System-on-Chips". CMP media LLC, Embedded Systems Conference, April 2003, San Francisco.

[6] C. MacNamee, and D. Heffernan, "Emerging on-chip debugging techniques for real-time embedded systems", IEE Computing & Control Engineering Journal, vol. 11, no. 6, Dec. 2000, pp. 295-303.

[7] K. Lemon, "Introduction to the Universal Measurement and Calibration Protocol XCP", SAE World Congress, March 2003, Detroit, Michigan.